# Annular Channel Eigenmodes: A Physical-Layer Approach to Suppressing OAM Modal Crosstalk

**Chenghao Li,[1#] Pengyang An,[1,2#] Ziao Huang,[1] Qiwen Zhan,[3,4*] and Guanghui Yuan[1,2**]**

[1]*Department of Optics and Optical Engineering, School of Physical Sciences, University of Science and Technology of China, Hefei, Anhui 230026, China*

[2]*State Key Laboratory of Opto-Electronic Information Acquisition and Protection Technology, Hefei, Anhui 230601, China*

[3]*School of Optical-Electrical and Computer Engineering, University of Shanghai for Science and Technology, Shanghai, 200093, China*

[4]*Zhejiang Key Laboratory of 3D Micro/Nano Fabrication and Characterization, Department of Electronic and Information Engineering, School of Engineering, Westlake University, Hangzhou 310030, China*

[#]*Authors with equal contribution*

*\*qwzhan@usst.edu.cn*

*\*\*ghyuan@ustc.edu.cn*

**Abstract:** Modal crosstalk is a fundamental limitation for orbital angular momentum (OAM)-based spatial-division multiplexing. Here, we introduce Annular Channel Eigenmodes (ACEs) —rigorously derived as the optimal band-limited solution for maximizing energy concentration within distinct annular channels. This approach reformulates the design as a Hermitian eigenvalue problem, efficiently yielding optimal beams that are physically isolated in space. Numerical simulations demonstrate that under identical conditions, conventional Gaussian-enveloped perfect optical vortices (POVs) exhibit an average modal crosstalk of -16 dB, whereas ACEs suppress crosstalk to nearly -30 dB. Moreover, the crosstalk suppression of ACEs continues to improve exponentially with increasing channel width, while that of POVs saturates at a fundamental limit. We experimentally generated ACEs and confirmed a 36% enhancement in energy confinement relative to POVs. ACEs thus provide a physically robust basis for high-fidelity, high-density OAM communications.

## 1. Introduction

Precise light focusing has long been a cornerstone of optical research. As scientific inquiry advances into increasingly fine scales, the scope of optical manipulation continues to expand. A striking example is the orbital angular momentum (OAM) of light, whose unique phase structure and propagation characteristics have profoundly impacted diverse fields, including quantum science [1], optical communications [2], and super-resolution microscopy [3]. Yet, despite the rapid growth of OAM applications, research into its fundamental focusing behavior remains surprisingly limited. For conventional light fields, researchers have developed a wide range of advanced techniques—from complex structured light [4] and super-oscillations [5] to spatio-temporal control [6]—to achieve enhanced, multi-dimensional focusing. In stark contrast, the vast majority of OAM studies still rely on Laguerre-Gaussian (LG) modes as a default basis [7, 8].

While LG beams with distinct OAM values are theoretically orthogonal and separable via mode decomposition, their significant spatial overlap renders this orthogonality fragile. In practical systems, even minimal diffraction can severely disrupt this property—a challenge that has significantly hindered progress and practical implementation in related fields.

A more robust approach to OAM multiplexing involves generating spatially distinct annular beams [9]. A prominent example is the perfect optical vortex (POV), produced by Fourier transforming a truncated Bessel beam to form a ring of a specific radius [10]. However, this spectral truncation is mathematically equivalent to applying a hard-edged window, which inevitably induces the Gibbs phenomenon [11]. Consequently, POVs are inherently plagued by strong sidelobes that degrade beam quality and interfere with adjacent channels (Fig. 1, bottom row) [12]. While methods utilizing a Gaussian envelope have been proposed to mitigate these sidelobes [13, 14], this work will demonstrate that it is possible to achieve far superior spatial separation than even POVs specifically optimized with a Gaussian envelope, manifested as a significantly higher signal-to-noise ratio (SNR) between channels.

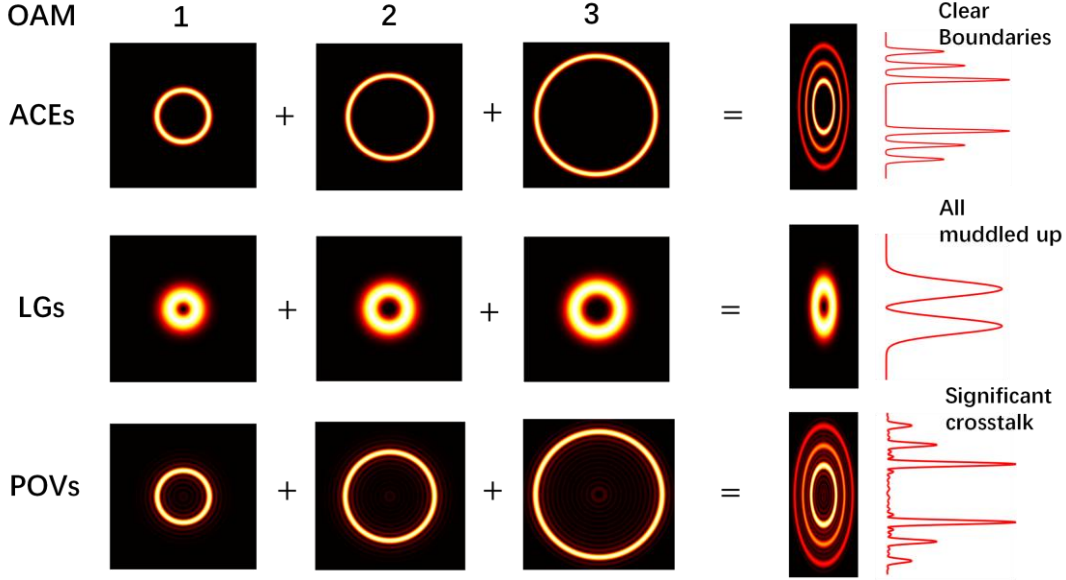

Figure 1. Comparison of Annular Channel Eigenmodes (ACEs) with common vortex beam modes. From top to bottom: ACEs, Laguerre-Gaussian (LG) modes, and Perfect Optical Vortices (POVs). Left of the equal sign: simulated single-ring profiles of each mode under identical imaging conditions. Right of the equal sign: simulated superposition profiles of the rings for each mode. Far right: corresponding radial intensity distributions of the superposed fields.

In this paper, we introduce a new class of OAM light fields, termed Annular Channel Eigenmodes (ACEs), which are rigorously derived as the optimal solutions for maximizing the energy concentration of a band-limited optical field within a predefined annular channel $[r_a, r_b]$ in the focal plane. Our formulation inherently incorporates the finite pupil constraint, transforming this complex optimization task into a standard matrix eigenvalue problem that can be solved efficiently [15]. The largest eigenvalue of the corresponding Hermitian operator represents the physical limit of energy efficiency, while its associated eigenvector yields the optimal pupil function required to attain this limit.

Crucially, ACEs offer a comprehensive solution to the primary mode separation challenges faced by both LG beams and POVs. First, ACEs are intrinsically free from strong sidelobes. Unlike the hard-clipped spectrum of a POV, the optimal pupil function of an ACE is naturally apodized within the finite pupil, tapering smoothly to zero at the pupil edge. This fundamentally suppresses the Gibbs phenomenon, resulting in remarkably clean annular profiles, as shown in Fig. 1 (top row). Second, ACEs provide far greater flexibility in beam engineering. While a POV is rigidly defined by a single target radius, an ACE can be optimized for an annular channel of any desired width and position, enabling precise, custom-shaped ring fields. Finally, ACEs establish a more robust foundation for orthogonality. By designing modes for spatially distinct, non-overlapping annular channels, their orthogonality is physically enforced via spatial separation rather than relying on a fragile mathematical condition in an overlapping domain, making them highly robust for practical mode-division multiplexing [16]. As numerical simulations later demonstrate, crosstalk between adjacent ACEs modes can drop below -100 dB and continues to improve as channel width increases, laying a solid foundation for spatial mode multiplexing.

Building upon these properties, ACEs exhibit strong potential for a range of applications where precise spatial confinement and mode orthogonality are critical. In optical manipulation, their clean and tunable annular intensity distributions enable the creation of stable ring-shaped traps and controlled multi-particle configurations [17, 18]. In quantum state discrimination, they can serve as spatially orthogonal detection bases with minimal modal overlap, improving the fidelity of high-dimensional quantum measurements [19, 20]. In free-space optical communications, the sharply confined annular channels of ACEs support dense, low-crosstalk OAM multiplexing with potential robustness against atmospheric turbulence [21–23]. These capabilities highlight that ACEs provide a versatile and physically robust framework for next-generation optical systems requiring both high spatial precision and reliable mode separation.

## 2. Methods and Numerical Results

In this section, we detail the numerical method developed to solve the problem of maximizing the energy concentration within a predefined annular channel for a band-limited optical system. As schematically illustrated in Fig. 2(a), the propagation of a light field from the pupil plane to the focal plane of an aberration-free optical system is modeled as a two-dimensional Fourier transform. The pupil function, defined within a finite circular aperture, represents the band-limited nature of the system.

In the application scenarios considered, such as mode-division multiplexing, the light field is characterized by its orbital angular momentum (OAM). It is therefore advantageous to express the 2D Fourier transform in polar coordinates. For a pupil function separable in amplitude and phase, $P(\rho,\phi) = A(\rho)e^{i\ell\phi}$, its corresponding field in the focal plane, $U(r,\theta)$, is given by:

$$U(r,\theta) = C \int_0^{a_{\text{pupil}}} \int_0^{2\pi} A(\rho)e^{i\ell\phi}e^{i\frac{2\pi NA}{\lambda}\rho r\cos(\phi-\theta)}\rho\,\mathrm{d}\rho\,\mathrm{d}\phi \qquad (1)$$

where $C$ is a complex constant, $A(\rho)$ is the radial amplitude profile in the pupil plane, $a_{\text{pupil}}$ is the pupil radius, $NA$ is the numerical aperture of the system, and $\lambda$ is the wavelength.

The angular integration reveals that the transformation of the radial component $A(\rho)$ corresponds to an $\ell$-th order Hankel transform, therefore, equation (1) can be rewritten as:

$$U(r,\theta) = C \cdot e^{i\ell\theta} \int_0^{a_{\text{pupil}}} A(\rho) J_\ell\left(\frac{2\pi NA}{\lambda}\rho r\right)\rho\,\mathrm{d}\rho \qquad (2)$$

where $J_\ell(r)$ is the $\ell$-th order Bessel function of the first kind. In our numerical framework, this continuous integral transform is discretized into a matrix $H_\ell$, which operates on a vector representing the radial pupil function $A(\rho)$.

To minimize inter-modal crosstalk, spatial separation of channels is required. This is achieved by confining the energy distribution of each mode to a specific, non-overlapping annular domain. As illustrated in Fig. 2(a), this constraint is implemented via an annular virtual aperture defined in the focal plane. The optimization objective thus becomes maximizing the ratio of the energy contained within this annulus to the total beam energy. This optimization problem is equivalent to finding the largest eigenvalue of a specific Hermitian operator.

To solve this problem numerically, we first discretize the continuous fields and operators. In the spectral domain (pupil plane), the field is physically band-limited by a circular aperture with $k_0 = 2\pi \cdot$

$NA/\lambda$. Leveraging the field's axial symmetry, we represent the radial amplitude function as a discrete column vector $P$ of size $N_\rho$. The components of $P$ are defined as $P_j = A(\rho_j) \cdot \sqrt{\rho_j w_j}$, where $A(\rho_j)$ is the function value at the $j$-th Gaussian-Legendre node, and $w_j$ is the corresponding integration weight (used to enhance numerical accuracy).

The field in the space domain (focal plane) is, in theory, of infinite extent. However, for our optimization, only the field within a finite computational domain that fully encompasses the target annulus $[r_a, r_b]$ needs to be considered. The continuous focal-plane field $U(r)$ is thus discretized into a vector $U$ of size $N_r$ with components $U_k = E(r_k)$. To mitigate potential boundary errors, the computational range for $r_k$ in this work was set to $[0, 1.5r_b]$. The linear transformation between the pupil and focal planes corresponds to the discrete Hankel transform, represented by the matrix $H_{kj} = J_\ell(k_0 \rho_j r_k)\sqrt{\rho_j w_j}$.

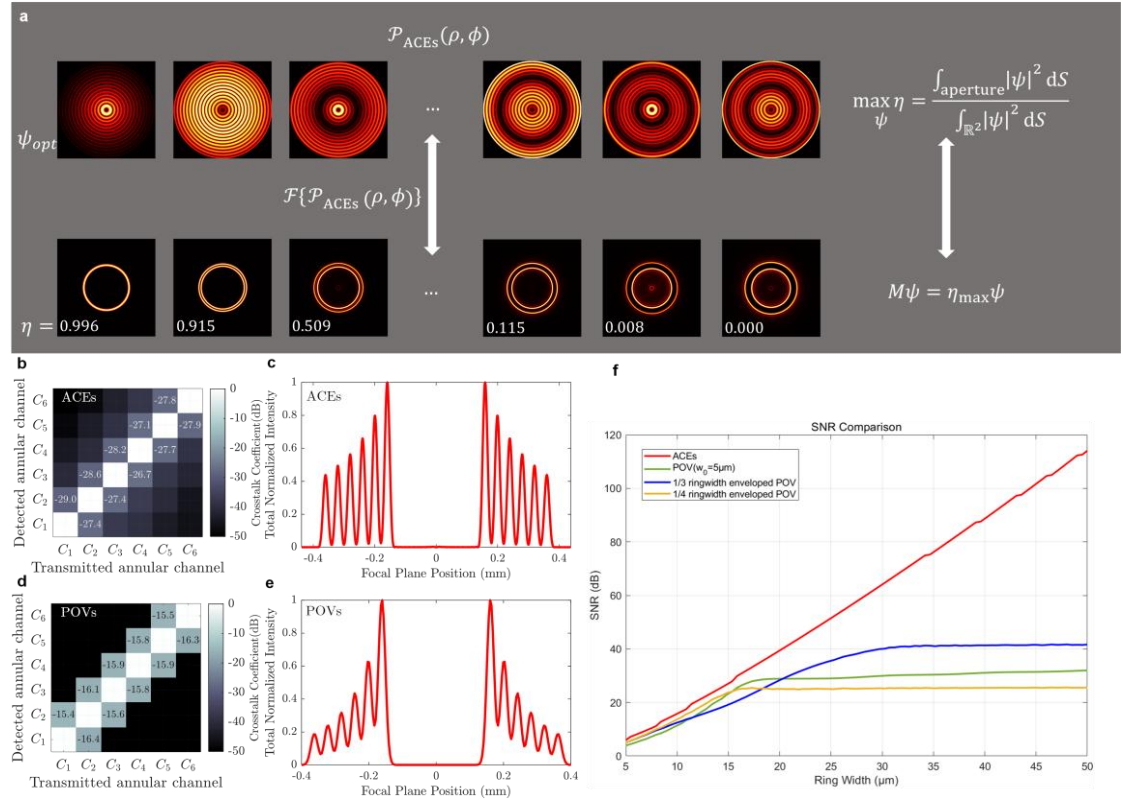

Figure 2. Principle and numerical results of ACEs. (a) Conceptual diagram of the ACEs design principle: the upper panel corresponds to the focal plane, and the lower panel corresponds to the pupil (spectral) plane. $\psi_{opt}$ represents the optimal eigenmode of the eigenvalue equation in the pupil plane, while the eigenvalue $\eta$ denotes the annular energy concentration efficiency. (b) Simulated crosstalk matrix for ACEs. (c) Simulated radial intensity distribution of ACEs in the focal plane. (d) Simulated crosstalk matrix for POVs. (e) Simulated radial intensity distribution of POVs in the focal plane. (f) Energy concentration efficiency of different modes as a function of the target annular width. Noise is defined as the total energy spilled outside the target annulus. Simulation parameters in (b)-(e) are consistent with experiment parameters. Simulation parameters in (f) are set to $NA$ = 0.05 and a central ring radius of 0.1 mm for generality.

Furthermore, according to Parseval's theorem, the total beam energy can be calculated accurately in the pupil plane. Due to our definition of the vector $P$, the total energy simplifies to the squared norm of

$P$, given by $E_{total} = P^\dagger P$. To calculate the energy within the target annulus, we define an annular selection operator $S$:

$$S_{kk} = \begin{cases} r_k \cdot dr & \text{if } r_a \leq r_k \leq r_b \\ 0 & \text{otherwise} \end{cases} \quad (3)$$

The energy contained within the target annulus, $E_{annal}$, is derived from the pupil-plane state vector as $E_{annulus} = P^\dagger H^\dagger SHP$. Consequently, the energy concentration ratio, which we aim to maximize, is expressed as the Rayleigh quotient for the operator $M = H^\dagger SH$:

$$\eta = \frac{P^\dagger H^\dagger SHP}{P^\dagger P} \quad (4)$$

A well-established principle in linear algebra states that the maximum value of this Rayleigh quotient corresponds to the largest eigenvalue of $M$. Thus, maximizing the energy ratio $\eta$ reduces to solving the standard eigenvalue problem $MP = \eta P$.

The maximum achievable energy concentration, $\eta_{max}$, is the largest eigenvalue of $M$, and the corresponding eigenvector, $P_{opt}$, represents the optimal pupil function. This matrix eigenvalue problem can be solved efficiently and accurately using standard numerical linear algebra routines.

In Figs. 2(b)-2(f), we numerically demonstrate the superiority of ACEs over Gaussian-enveloped POVs for spatial energy separation, highlighting their significant potential for spatial-division multiplexing in optical communications. We first calculate the focal-plane intensity distributions of both ACEs and POVs under identical system parameters with the experiential setup ($NA = 0.0175, \lambda = 532$ nm). Six equidistant annular channels are defined within the radial range of 140 μm to 380 μm. For the POVs basis, the ring radius of each channel is set to the center of the corresponding ACEs channel, and the Gaussian envelope waist in the pupil plane, $w_g$, is chosen to yield a focal-plane ring width equal to half the channel width (40 μm), ensuring strong sidelobe surpression. The simulated intensity profiles, shown in Figs. 2(a) and 2(c), reveal that the ACEs modes exhibit remarkably smooth profiles with sharp boundaries. In contrast, the POVs modes are plagued by dense crosstalk—an inherent limitation that constrains the achievable SNR. This is quantitatively confirmed by the crosstalk matrices in Figs. 2(b) and 2(d): the adjacent-channel crosstalk for POVs is approximately -16 dB, whereas for ACEs it is suppressed to reach -30 dB. Furthermore, Fig. 2(f) illustrates this advantage more comprehensively. The SNR for the ACEs basis is consistently higher than that of any POV configuration. As channel width increases, the SNR of ACEs continues to improve, while the SNR of POVs saturates, revealing a fundamental performance ceiling. Based on our numerical simulations, the optimal SNR for POVs is generally achieved when the pupil-plane waist $w_g$ is chosen to produce a focal-plane ring width between 1/3 and 1/4 of the channel width. This represents a trade-off between maximal sidelobe suppression (wider ring) and tightest beam confinement (narrower ring). Curves for other parameter choices will fall between these two representative cases.

## 3. Experimental Results

To validate the performance of ACEs in suppressing OAM modal crosstalk, we established the optical characterization system illustrated in Fig. 3(a). A laser beam with a wavelength of $\lambda = 532$ nm was expanded, collimated, and then incident on the spatial light modulator (SLM). Using the chessboard encoding method [24], the optimized pupil function $\mathcal{P}_{ACEs}(\rho, \phi)$, obtained via numerical calculation for different annular channels $C_i$, was loaded onto the SLM to shape the wavefront of the incident Gaussian beam. The shaped optical field was then Fourier-transformed by a lens (L1, focal length $f_1 = 250$ mm), generating the ACEs field at its focal plane (equivalent $NA = 0.0175$). The target field is filtered out from the first-order diffraction using the spatial filter. This field was subsequently imaged onto a high-resolution CCD camera via a 4f imaging system (L2, L3) to record the focal-plane intensity distribution $I(r,\theta) = |U(r,\theta)|^2$. Figs. 3(b) and 3(c) show the pupil function of $C_3$ (amplitude and phase) and the corresponding simulated ACEs field (intensity and phase), respectively. Fig. 3(d) presents the pure phase map of the pupil function of $C_3$ generated after checkerboard encoding, which was subsequently loaded onto the SLM to modulate the incident light beam.

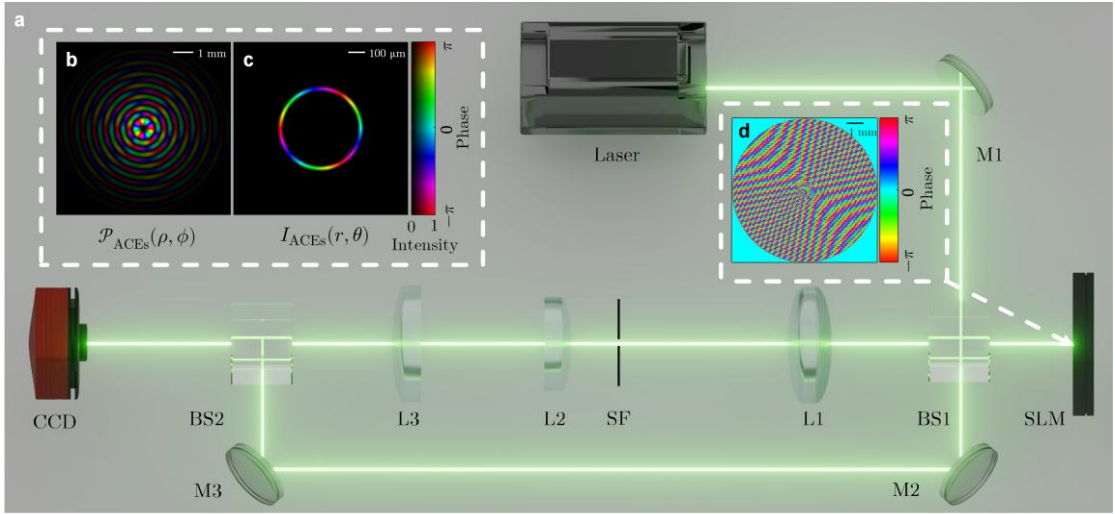

Figure 3. Experimental setup for generating and characterizing ACEs. (a) Schematic of the optical characterization system. M, mirror; BS, beam splitter; SLM, spatial light modulator; SF, spatial filter; L, lens; CCD, charge-coupled device. (b) Exemplary pupil function (amplitude and phase) for annular channel $C_3$. (c) Corresponding simulated ACEs field (intensity and phase) for the pupil function in (b). (d) SLM phase map implementing the pupil function in (b) using the checkerboard method.

For comparison, Gaussian-enveloped POVs were generated using the same optical path. Their pupil function is given by $\mathcal{P}_{\text{POVs}}(\rho,\phi) = J_\ell(k_\rho \rho)\exp(i\ell\phi)\exp(-\rho^2/\omega_g^2)$, where parameters $k_\rho$ and $\omega_g$ were set such that the target ring radius $r_r = k_\rho \cdot f_1/k$ and ring width matched those of the corresponding ACEs channel for a fair comparison [14]. Additionally, a Gaussian beam split off by beam splitter BS1 served as a reference for subsequent verification of topological charge via interferometry.

We first characterized the focal-plane optical fields of the ACEs modes across six annular channels ($C_1 - C_6$). As shown in Fig. 4(a), the experimentally measured intensity and phase distributions of the ACEs are strikingly consistent with the numerical simulations, exhibiting clean annular structures that are strictly confined within the designed channels with significantly suppressed sidelobes. This demonstrates exceptional spatial separation capability of ACEs.

For quantitative comparison, Fig. 4(b) plots the radially averaged intensity profiles for both ACEs and POVs across all six channels. It is clearly observed that the intensity profile of the ACEs exhibits a higher peak within the target channel and a much steeper decay beyond the channel boundaries, whereas the POVs show pronounced broadening and energy leakage. This directly confirms the superior energy concentration of the ACEs.

Modal crosstalk is a key metric for evaluating the performance of mode-division multiplexing systems. We constructed the crosstalk matrix from experimental measurements, where its elements are defined as $XT_{ji} = P_{ji}/P_{ii}$, with $P_{ji}$ being the power detected in channel $C_j$ when channel $C_i$ is excited. Figs. 4(c) and 4(d) present the experimental crosstalk matrices for ACEs and POVs, respectively. For the ACEs, the average off-diagonal crosstalk is suppressed below -13 dB, whereas for the POVs, the corresponding value is only -11 dB. This distinct 2 dB improvement demonstrates the enhanced channel isolation achieved by ACEs at the physical layer.

Finally, we analyzed the power concentration of each mode within their respective target channels, defined as $\eta = E_{obj}/E_{total}$, where $E_{obj}$ is the energy within the target annulus and $E_{total}$ is the total energy. The results are shown in Fig. 4(e). The power concentration for all ACEs modes exceeds 90% (indicated by the grey dashed line), while the average for the POVs modes is only 86.8%, indicating a 36% enhancement in energy confinement capability (defined as the inverse of the energy spillover ratio). This finding further solidifies the superiority of ACEs as high-performance spatial channel modes from the perspective of energy efficiency.

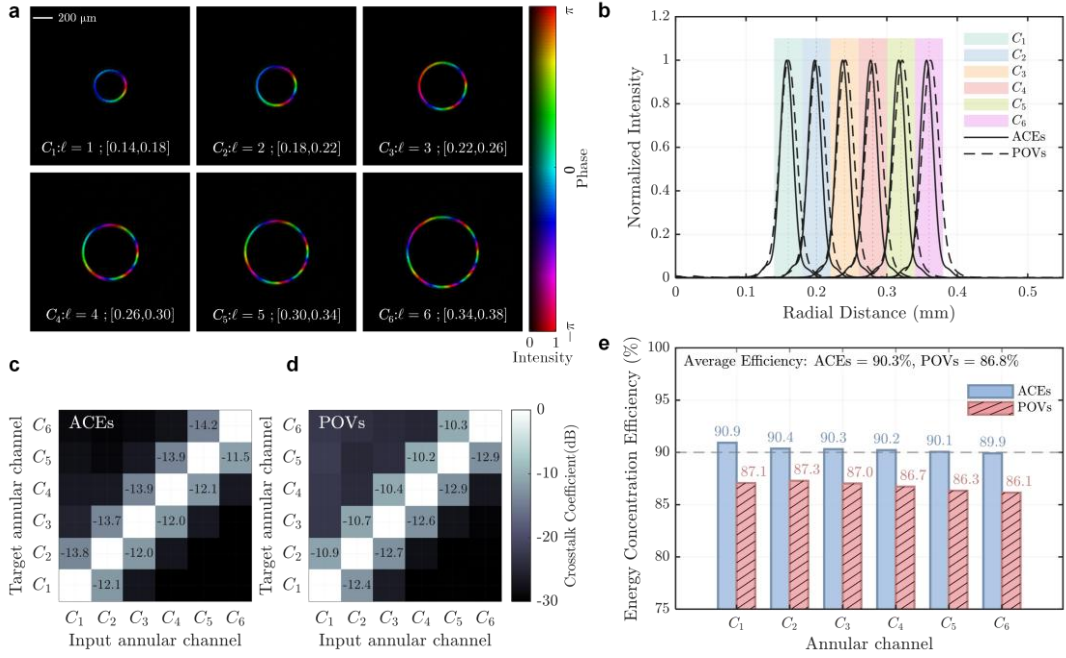

Figure 4. Experimental performance comparison between ACEs and POVs. (a) Experimentally measured focal-plane intensity and phase distributions of the ACEs across six annular channels ($C_1$-$C_6$). (b) Comparison of experimental radial intensity profiles between ACEs and POVs for all six channels. Colored shaded areas indicate the designed extent of each channel. (c), (d) Experimentally constructed crosstalk metrics: (c) ACEs; (d) POVs. ACEs exhibit lower off-diagonal crosstalk. (e) Statistics of power concentration within respective target channels for ACEs and POVs. The grey dashed line indicates the 90% benchmark.

## 4. Conclusion

In summary, we have introduced and experimentally demonstrated annular channel eigenmodes (ACEs)—a new class of OAM light fields rigorously derived as the optimal solutions for maximizing spatial energy confinement within predefined annular channels. By formulating the design problem as a Hermitian eigenvalue problem, ACEs inherently achieve smooth pupil apodization and effectively suppress sidelobes, overcoming the fundamental limitations of both Laguerre–Gaussian beams and perfect optical vortices. Experimental results confirm that ACEs exhibit significantly enhanced energy concentration and reduced inter-channel crosstalk, providing a physically robust basis for spatial mode separation. These properties make ACEs a powerful platform for next-generation optical systems, particularly in optical manipulation, quantum state discrimination, and free-space optical communications, where high spatial precision and mode orthogonality are essential.

**Acknowledgement**

This work was supported by the CAS Project for Young Scientists in Basic Research (Grant No. YSBR-049) and the Overseas Excellent Youth Science Foundation Project. Qiwen Zhan also acknowledge financial support from National Natural Science Foundation of China (NSFC) (Grant No. 12434012). Pengyang An and Chenghao Li would like to thank Chengda Song for the fruitful discussions and insightful suggestions, which have provided valuable support for experimental design and code optimization.

**Disclosures**

The authors declare no conflicts of interest.

**Data availability**

Codes and data underlying the results presented in this paper are available from the corresponding authors, upon reasonable request.